\documentclass[letterpaper]{article}

\pdfoutput=1

\usepackage[preprint]{neurips_2021}
\usepackage[utf8]{inputenc} 
\usepackage[T1]{fontenc}    
\usepackage[hyperindex,breaklinks]{hyperref}
\usepackage{url}   
\usepackage{microtype}      

\setcitestyle{authoryear,open={(},close={)}}

\title{Addressing Privacy Threats from Machine Learning}

\author{%
  Mary Anne Smart \\
  Department of Computer Science \& Engineering\\
  University of California San Diego\\
  \texttt{msmart@ucsd.edu} \\
}

\begin{document}

\maketitle

\begin{abstract}
  Every year at NeurIPS, machine learning researchers gather and discuss exciting applications of machine learning in areas such as public health, disaster response, climate change, education, and more\footnote{These topics are all the subjects of \href{https://blog.neurips.cc/2021/08/20/neurips-2021-workshop-announcement/}{NeurIPS 2021 workshops}.}. However, many of these same researchers are expressing growing concern about applications of machine learning for surveillance \citep{BroaderImpact}. This paper presents a brief overview of strategies for resisting these surveillance technologies and calls for greater collaboration between machine learning and human-computer interaction researchers to address the threats that these technologies pose.
\end{abstract}

\section{Introduction}

A rapidly growing research area in computer science is that of privacy in machine learning \citep{Papernot2018, Mirshghallah2020, Liu2021}. Differentially-private machine learning algorithms and other privacy-preserving methods offer the opportunity to learn from sensitive datasets while limiting what can be revealed about any particular individual. However, some machine learning systems violate privacy---not as some unintended side-effect, but by design. The tools from privacy-preserving machine learning are little use against machine learning models that are designed for biometric or behavioral profiling \citep{Ligett2020}; different approaches are needed.

Although a variety of definitions, frameworks, and taxonomies have been proposed, there is no single, universally agreed-upon definition of privacy \citep{Solove2008, Nissenbaum2009, Arora2019, McDonald2020}. This paper focuses on the specific set of privacy-related harms that are perpetrated or exacerbated by machine learning systems. In an age where powerful algorithms can process large datasets at high speeds, it becomes harder to find privacy by hiding in a crowd. Facial recognition algorithms can identify people in public spaces \citep{Garvie2016, Stark2019}; targeted advertising algorithms can exploit detailed user profiles to shape behavior at scale \citep{Zuboff2018}; and predictive policing algorithms can single out individuals for targeted surveillance \citep{Stroud2021}. These are the kinds of novel threats to privacy that machine learning systems enable. 

A number of strategies have emerged for resisting machine learning systems that threaten privacy. This work provides an overview of such strategies and argues that addressing these privacy challenges effectively will require greater collaboration between the NeurIPS community and experts in the field of human-computer interaction (HCI). This paper discusses two overarching approaches for addressing threats to privacy posed by machine learning (although in some cases the line between the two may be blurred). The first approach involves challenging the data that feeds the malicious machine learning model, either through some form of obfuscation or through withholding information. The second approach involves challenging the malicious model directly, perhaps by applying pressure to force the retirement of the model or by banning the harmful technology altogether. With either approach, there is a role for computer scientists to play.

\section{Challenging Data}
Machine learning systems run on data. First, data is required to train a machine learning model. Next, after the model has been deployed, new data is fed into the model to produce predictions. In practice, these stages of training and deployment may take place iteratively; for example, the deployed model may periodically be retired and replaced by a new model retrained on the latest data. One way to resist a machine learning system is to interfere with the data that feeds it. There are two main strategies that fall under this umbrella: obfuscation and withholding.

\subsection{Obfuscation}

One way to evade machine learning surveillance systems is by strategically altering either 1) the data used to make predictions or 2) the data used to train the system. As an example of the first strategy, \cite{Sharif2016} designed glasses that could fool facial recognition systems. The researchers exploited the fact that machine learning models tend to be vulnerable to \textit{adversarial examples}. An adversarial example is a datapoint that has been modified slightly--such that a human would not perceive any difference but that a machine learning model misclassifies it. In recent years, researchers have developed a number of strategies for evading facial recognition using adversarial examples \citep{Harvey2017, Thys2019, CherepanovaGFDD21, Li2021, Chandrasekaran2021}. These approaches aim to help individuals avoid surveillance but typically lack strong guarantees \citep{Rajabi2021, Bauer2021, Chen2021, Radiya2021}.

An alternative to altering data at the time of prediction is to alter the data used to train the underlying machine learning model. This style of attack is known as a \textit{data poisoning attack}. For example, TensorClog produces altered images that are designed to reduced the accuracy of deep learning models \citep{Shen2019}. Another example of data poisoning for anti-surveillance purposes can be found at Adversarial Fashion\footnote{\url{https://adversarialfashion.com/}}, a shop that sells clothing designed to trigger automated license plate readers and "[inject] junk data" \citep{Rose2021}. 

Although the above examples focus on image classification tasks, similar obfuscation tactics have been used for resisting web tracking, loyalty card-based tracking, and more \citep{brunton2015obfuscation, pots}. While obfuscation can be of practical utility, it can also have an \textit{expressive} function \citep{nissenbaum2009trackmenot, Howe2015surveillance}. For example, members of the "Dazzle Club" paint their faces with unusual makeup and conduct walks through the streets of London \citep{dazzleclub}. Although their makeup is designed to trick facial recognition systems, the group's real purpose is to protest police use of facial recognition. 

Adversarial examples and data poisoning attacks are ongoing topics of study within the machine learning community. If these technologies are to be adopted as anti-surveillance tools, many practical and ethical challenges remain \citep{brunton2015obfuscation, ObfuscationWorkshop2017, das2020subversive, Albert2020}. How can such tools be made more accessible? How should such tools be evaluated? How should the risks of using obfuscation be communicated? Addressing these questions will require collaborations among machine learning experts, HCI researchers, activists, and other stakeholders. 

\subsection{Withholding Data} 
An alternative to altering the data used by machine learning surveillance systems is to withhold the data. One simple example of this strategy is the use of privacy-enhancing technologies that block web tracking \citep{Mayer2012}. Withholding data by using tracker-blocking browser extensions may provide some privacy protection for individuals. However, data can also be withheld at the collective level \citep{vincent2021leverage}. For example, a \textit{data strike} is a kind of digital boycott that can be used to apply public pressure to technology companies \citep{vincent2019data}. A related way of withholding data is through \textit{protest non-use}; many people have simply stopped using certain platforms due to privacy and surveillance-related concerns \citep{li2019protest}. Dazzle club walks, data strikes, and protest non-use all go one step beyond strategies that obfuscate or withhold data solely as a way of evading surveillance; rather, they use these data-focused strategies as starting points to launch broader critiques and campaigns against the surveillance systems themselves.

\section{Challenging Models}
Although the data-oriented approaches described so far offer some help in resisting machine learning surveillance systems, in many cases, policy approaches offer a more satisfactory solution. For example, while a variety of strategies exist for evading or fooling facial recognition, banning particular uses of facial recognition would make these strategies unnecessary. There exist a wide range of forms that regulation can take and a wide range of roles that computer scientists may play in the process.

One technique that can apply pressure to companies that develop and sell surveillance technologies is \emph{auditing}. In the case of facial recognition, audits by researchers have helped call attention to the harm that this technology can cause. It is now well-established that many facial recognition systems perform poorly on darker-skinned subjects \citep{buolamwini2018gender, buolamwini2019actionable}; these accuracy issues have led to multiple wrongful arrests of Black men in the United States \citep{falsearrest}. The audits that uncovered racial bias in facial recognition systems placed pressure on companies that sell facial recognition technology to police. In fact, in June 2020, several major companies stopped selling facial recognition technology due to mounting pressure from researchers, protestors, and their own employees \citep{ibm2020}. Nevertheless, audits have limitations; for example, audits may inadvertently "normalize tasks that are inherently harmful to certain communities" \citep{raji2020}.

Some proprietary technologies cannot be easily audited because access to the technology is restricted.  Nevertheless, it is sometimes possible to reverse engineer a system and show that it can produce societal harm. For example, \cite{lum2016predict} simulate the effects of predictive policing systems and show how predictive policing can magnify existing biases and lead to the over-policing of low-income communities and communities of color. Many of the same risks involved with traditional audits are at play here as well. For example, a narrow focus on particular metrics could simply result in shifting goal posts -- what \cite{polack2020beyond} refers to as \textit{algorithmic reformism}. Therefore, when conducting algorithmic audits or when reverse engineering particular algorithms, it is important to consider the scope of the critique one hopes to make.

In some cases, researchers have partnered with community organizations in order to push back against surveillance technologies. Sometimes mathematical or technical language is used to insulate surveillance technologies from public criticism \citep{stop2014spying}. For example, an LAPD spokesperson infamously said of predictive policing: "It is math, not magic, and it is not racist" \citep{racistmath}. Technologists can debunk the myth that critics of surveillance technologies simply don't understand the math; they can also help parse technical language and demystify machine learning algorithms. However, it is important that researchers approach partnerships with humility; community organizers will bring their own areas of expertise to the table \citep{whitney2021hci}.

Finally, it is important to acknowledge the academic community's complicity in building and upholding the very forms of surveillance discussed in this paper \citep{stop2014spying}. \cite{Ko2020} argue that computer science educators have a responsibility to make the role of computing in injustice visible. At some universities, students have taken it upon themselves to educate each other about the harmful role of computing within US Immigration and Customs Enforcement\footnote{\url{https://notechforice.com}} \citep{Zong}. These kinds of efforts will help the next generation of computer scientists think critically about the consequences of the technologies they create.

\section{Conclusion}
This paper has outlined a range of strategies for resisting surveillance technologies powered by machine learning. It concludes by echoing \cite{das2020subversive} in calling for participatory methods when designing anti-surveillance technologies such as those discussed in this paper. While participatory methods are relatively common in HCI research \citep{Muller2012}, they have received less attention from the machine learning community\footnote{Some exceptions include the \href{https://participatoryml.github.io/}{Participatory Approaches to Machine Learning ICML 2020 Workshop} and the \href{https://sites.google.com/view/resistance-ai-neurips-20/home}{Resistance AI NeurIPS 2020 Workshop}.}. The impact of data-driven surveillance technologies is not borne equally by all; rather, the brunt of these technologies is disproportionately borne by already marginalized people \citep{marwick2018privacy, Eubanks2018, Roberts2019digitizing}. Although not a panacea \citep{Sloan2020participation}, participatory approaches can help ensure that the design of anti-surveillance technologies is led by those who are disproportionately targeted by surveillance. 

\bibliographystyle{abbrvnat}
\bibliography{references}

\end{document}